\documentclass[12pt]{article}
\usepackage{fancyhdr}

\usepackage{mathrsfs}
\usepackage[T1]{fontenc}
\usepackage{mathpazo}
\usepackage{setspace}
\usepackage{amsfonts}
\usepackage{amssymb}
\usepackage{epsfig}
\usepackage{latexsym}
\usepackage{color}
\usepackage{graphicx}
\usepackage{nicefrac}
\usepackage[latin1]{inputenc}
\usepackage{slashed}
\usepackage{multirow}

\usepackage{hyperref}

\def\hybrid{\topmargin -20pt    \oddsidemargin 0pt
        \headheight 0pt \headsep 0pt
        \textwidth 6.25in       
        \textheight 9.25in       
        \marginparwidth .875in
        \parskip 5pt plus 1pt   \jot = 1.5ex}

\hybrid

\def\baselinestretch{1.2}

\catcode`\@=11

\def\marginnote#1{}
\newcount\hour
\newcount\minute
\newtoks\amorpm
\hour=\time\divide\hour by60
\minute=\time{\multiply\hour by60 \global\advance\minute by-\hour}
\edef\standardtime{{\ifnum\hour<12 \global\amorpm={am}%
        \else\global\amorpm={pm}\advance\hour by-12 \fi
        \ifnum\hour=0 \hour=12 \fi
        \number\hour:\ifnum\minute<10 0\fi\number\minute\the\amorpm}}
\edef\militarytime{\number\hour:\ifnum\minute<10 0\fi\number\minute}

\def\draftlabel#1{{\@bsphack\if@filesw {\let\thepage\relax
   \xdef\@gtempa{\write\@auxout{\string
      \newlabel{#1}{{\@currentlabel}{\thepage}}}}}\@gtempa
   \if@nobreak \ifvmode\nobreak\fi\fi\fi\@esphack}
        \gdef\@eqnlabel{#1}}
\def\@eqnlabel{}
\def\@vacuum{}
\def\draftmarginnote#1{\marginpar{\raggedright\scriptsize\tt#1}}

\def\draft{\oddsidemargin -.5truein
        \def\@oddfoot{\sl preliminary draft \hfil
        \rm\thepage\hfil\sl\today\quad\militarytime}
        \let\@evenfoot\@oddfoot \overfullrule 3pt
        \let\label=\draftlabel
        \let\marginnote=\draftmarginnote
   \def\@eqnnum{(\theequation)\rlap{\kern\marginparsep\tt\@eqnlabel}%
\global\let\@eqnlabel\@vacuum}  }

\def\preprint{\twocolumn\sloppy\flushbottom\parindent 2em
        \leftmargini 2em\leftmarginv .5em\leftmarginvi .5em
        \oddsidemargin -.5in    \evensidemargin -.5in
        \columnsep .4in \footheight 0pt
        \textwidth 10.in        \topmargin  -.4in
        \headheight 12pt \topskip .4in
        \textheight 6.9in \footskip 0pt
        \def\@oddhead{\thepage\hfil\addtocounter{page}{1}\thepage}
        \let\@evenhead\@oddhead \def\@oddfoot{} \def\@evenfoot{} }

\def\numberbysection{\@addtoreset{equation}{section}
        \def\theequation{\thesection.\arabic{equation}}}

\def\underline#1{\relax\ifmmode\@@underline#1\else
        $\@@underline{\hbox{#1}}$\relax\fi}

\def\titlepage{\@restonecolfalse\if@twocolumn\@restonecoltrue\onecolumn
     \else \newpage \fi \thispagestyle{empty}\c@page\z@
        \def\thefootnote{\fnsymbol{footnote}} }

\def\endtitlepage{\if@restonecol\twocolumn \else \newpage \fi
        \def\thefootnote{\arabic{footnote}}
        \setcounter{footnote}{0}}  

\catcode`@=12
\relax

\def\figcap{\section*{Figure Captions\markboth
        {FIGURECAPTIONS}{FIGURECAPTIONS}}\list
        {Figure \arabic{enumi}:\hfill}{\settowidth\labelwidth{Figure
999:}
        \leftmargin\labelwidth
        \advance\leftmargin\labelsep\usecounter{enumi}}}
 \relax
\def\tablecap{\section*{Table Captions\markboth
        {TABLECAPTIONS}{TABLECAPTIONS}}\list
        {Table \arabic{enumi}:\hfill}{\settowidth\labelwidth{Table
999:}
        \leftmargin\labelwidth
        \advance\leftmargin\labelsep\usecounter{enumi}}}
 \relax
\def\reflist{\section*{References\markboth
        {REFLIST}{REFLIST}}\list
        {[\arabic{enumi}]\hfill}{\settowidth\labelwidth{[999]}
        \leftmargin\labelwidth
        \advance\leftmargin\labelsep\usecounter{enumi}}}
 \relax
\makeatletter
\newcounter{pubctr}
\def\publist{\@ifnextchar[{\@publist}{\@@publist}}
\def\@publist[#1]{\list
        {[\arabic{pubctr}]\hfill}{\settowidth\labelwidth{[999]}
        \leftmargin\labelwidth
        \advance\leftmargin\labelsep
        \@nmbrlisttrue\def\@listctr{pubctr}
        \setcounter{pubctr}{#1}\addtocounter{pubctr}{-1}}}
\def\@@publist{\list
        {[\arabic{pubctr}]\hfill}{\settowidth\labelwidth{[999]}
        \leftmargin\labelwidth
        \advance\leftmargin\labelsep
        \@nmbrlisttrue\def\@listctr{pubctr}}}
 \relax
\makeatother
\newskip\humongous \humongous=0pt plus 1000pt minus 1000pt

\newif\ifdtup

\relax

\def\be{\begin{equation}}
\def\ee{\end{equation}}
\def\ba{\begin{eqnarray}}
\def\ea{\end{eqnarray}}

\def\del{\partial}

\def\k{\kappa}
\def\r{\rho}
\def\a{\alpha}

\def\b{\beta}

\def\g{\gamma}

\def\d{\delta}
\def\D{\Delta}

\def\m{\mu}
\def\n{\nu}
\def\om{\omega}
\def\Om{\Omega}
\def\l{\lambda}
\def\L{\Lambda}
\def\s{\sigma}

\def\cN{{\cal N}}

\def\cL{{\cal L}}

\def\no{\noindent}

\def\qq{\qquad}

\def\IR{\relax{\rm I\kern-.18em R}}

\def \ha {{1\over 2}}

\def \ov {\over}

\def\diag{{\rm diag}}

\def\IR{\relax{\rm I\kern-.18em R}}
\def\IL{\relax{\rm I\kern-.18em L}}

\def\inv{^{\raise.15ex\hbox{${\scriptscriptstyle -}$}\kern-.05em 1}}

\def\cL{{\cal L}}

\begin{document}

\renewcommand{\theequation}{\thesection.\arabic{equation}}
\csname @addtoreset\endcsname{equation}{section}

\newcommand{\beq}{\begin{equation}}
\newcommand{\eeq}[1]{\label{#1}\end{equation}}
\newcommand{\ber}{\begin{eqnarray}}
\newcommand{\eer}[1]{\label{#1}\end{eqnarray}}
\newcommand{\eqn}[1]{(\ref{#1})}
\begin{titlepage}
\begin{center}

\hfill DMUS--MP--13/23

\vskip  .7in

{\large \bf Integrable interpolations: From exact CFTs to non-Abelian T-duals}

\vskip 0.4in

{\bf Konstadinos Sfetsos}$^{1,2}$
\vskip 0.1in
{\em
${}^1$Department of Mathematics, University of Surrey,\\
Guildford GU2 7XH, UK\\
{\tt \footnotesize k.sfetsos@surrey.ac.uk}\\
\vskip .15in
${}^2$Department of Physics, 
University of Athens,\\
15771 Athens, Greece.\\
}

\vskip .5in
\end{center}

\centerline{\bf Abstract}

\no
We derive two new classes of integrable theories interpolating
between exact CFT WZW or gauged WZW models and non-Abelian T-duals of principal chiral models or geometric coset models.
They are naturally constructed by gauging symmetries of integrable models.
Our analysis implies that non-Abelian T-duality preserves integrability and
suggests a novel way to understand the global properties of the corresponding backgrounds.

\newpage

\tableofcontents

\noindent

\vskip .4in

\end{titlepage}
\vfill
\eject

\def\baselinestretch{1.2}
\baselineskip 20 pt
\noindent


\setcounter{equation}{0}
\section{Introduction and motivation}
\renewcommand{\theequation}{\thesection.\arabic{equation}}

Based on work in \cite{Rajeev:1988hq} it has been proposed that a way to construct integrable models
is to encode the conditions ensuring classical integrability into the Hamiltonian equations of motion following from general
current algebras with a underlying structure based on a group \cite{Balog:1993es}.
In principle this results into a generalization of the conformal currents algebras for the Wess--Zumino--Witten (WZW) model
as well as for the current algebras for the Principal Chiral Models (PCM) and Pseudochiral Models.
The authors of  \cite{Balog:1993es} obtained general condition that the $\s$-model fields have to satisfy in order for it to be integrable.
Moreover with a brute force computation they solved these conditions for the case that the group is $SU(2)$ and they explicitly constructed
a family of integrable models interpolating between the exact CFT WZW model with affine current algebra $SU(2)_k$
and the non-Abelian T-dual of the
PCM for $SU(2)$ with respect to $SU(2)_L$. The latter fact was explicitly pointed out in \cite{Evans:1994hi},
where in addition an $S$-matrix for this model was proposed and further checked.

\no
The above work was essentially not developed any further due to the fact that solving the general conditions of \cite{Balog:1993es}
seemed extremely difficult if not impossible given also the fact that in the simplest $SU(2)$ case the solution
was obtained by a brute force computation. Moreover, it was not even clear that there can be any other solutions.
In addition, any relation to other known constructions or of a systematic way to obtain solutions was lacking.
Hence, this way of constructing integrable models stood so far as an aesthetically appealing isolated example.

\no
The purpose of the present paper is to fill all of the above gaps and moreover to generalize the construction.
We will obtain integrable models for any group $G$ and explicit
expressions for the $\s$-model fields which are non-singular. These will interpolate between the exact CFT WZW model
with affine current algebra $G_k$ and the non-Abelian T-dual of the
PCM for $G$ with respect to the left action of the group denoted by $G_{\rm L}$.
More importantly, we will show that these models arise very naturally
by gauging the combined action for the PCM and the WZW for $G$.
The latter approach is inspired by a suggestion made in \cite{Sfetsos:1994vz} and further elaborated in \cite{Polychronakos:2010hd},
involving the relation of non-Abelian T-duality to gauged WZW models.
The deformation from the CFT point is driven by relevant operators of current bilinears.
In addition, we also suggest a new class of theories which interpolate between exact coset $G_k/H_k$ CFTs
realized using gauged WZW models and the non-Abelian T-duals of geometric cosets $G/H$ models.
In this case the deformation is driven by the parafermions of
the coset CFT. We check this assertion for the simplest case in which the group is $G=SU(2)$ and the subgroup is $H=U(1)$
making connection with existing work in \cite{Fateev:1991bv}. Therefore, we believe that these theories are also integrable.

\no
The original motivation for this work was to understand global issues
in non-Abelian T-duality \cite{Fridling:1983ha}-\cite{de la Ossa:1992vc} for which there has been a
considerable advancement in recent years, concerning in particular a complete understanding in type-II supergravity backgrounds
with non-trivial RR-fields \cite{Sfetsos:2010uq}-\cite{Itsios:2013wd} (and references therein).
In this direction we will make an important step since, for the case of only NS-sector fields, we will show
that models obtained via a non-Abelian T-duality transformation can be thought of as the end point of a line of integrable theories.

\no
The organization of this paper is as follows: In section 2 we first develop the general conditions
for integrability in relation to current algebras with structure dictated by group theory. Details on the
derivation of the corresponding Poisson brackets are given in Appendix A. Then we present the
solution to these conditions and the details of the proof in Appendix C.
In section 3 we present the origin of our integrable models by showing that they arise by gauging a subgroup of
the sum of a WZW model and a PCM for a non-semisimple group.
In section 4 we present an example based on $SU(2)$. In section 5 we generalize our construction for models with less symmetry
having also spectator fields inert under the action of the symmetry group. In section 6 we
construct theories interpolating between the exact coset $G_k/H_k$ CFT models and the non-Abelian T-duals of the $\s$-models
corresponding to the $G/H$ geometric coset models. We present evidence that these are also integrable. We end our paper with concluding
remarks in section 7. Besides Appendices A and C we also have written Appendix B where we collect some useful expressions needed in our
algebraic manipulations and Appendix D where an independent proof for the
integrability of the non-Abelian T-dual of the PCM model is given. Finally, in Appendix E we make a comparison of the
result of our construction for the simplest case of $SU(2)$ case with those of \cite{Balog:1993es}.

\setcounter{equation}{0}
\section{Setting up the models and integrability conditions}
\renewcommand{\theequation}{\thesection.\arabic{equation}}

We will consider a class of two-dimensional $\s$-models of the form
\be
S(X) ={1\ov 2} \int Q_{\m\n} \del_+ X^\m \del_- X^\n\ ,\qq Q_{\m\n}= G_{\m\n}+ B_{\m\n}\ ,
\label{eq1}
\ee
where $G_{\m\n}$ and $B_{\m\n}$ are the metric and the antisymmetric tensor fields, respectively.
The corresponding Lagrangian density can be written as\footnote{
We will use the coordinate conventions
\ba
&& x^{\pm } = x^0 \pm x^1 \ ,\qq x^0 = \ha (x^++x^-)\ ,\qq x^1 = \ha (x^+ - x^-)  \ ,
\nonumber\\
&& \del_\pm = \ha (\del_0\pm \del_1) \  ,\qq \del_0 =\del_+ + \del_- \ ,\qq \del_1 =\del_+ - \del_-\ .
\ea
We will also frequently denote $x^0$ and $x^1$ by $\tau$ and $\s$, respectively.
}
\ba
\cL & = & \ha (G_{\m\n} + B_{\m\n})\del_+ X^\m \del_- X^\n
\nonumber\\
&  = & {1\ov 8} G_{\m\n}( \del_0X^\m \del_0 X^\n - \del_1 X^\m \del_1 X^\n)
-{1\ov 4} B_{\m\n} \del_0 X^\m \del_1 X^\n\ .
\ea
The conjugate momentum to $X^\m$ is
\be
P_\m = {1\ov 4}\left(G_{\m\n} \del_0 X^\n - B_{\m\n} \del_1 X^\n\right)\
\ee
and the Hamiltonian is given by the integral $H= \int d\s {\cal H}$, where the density is
\be
{\cal H} = {1\ov 8} G_{\m\n} (\del_0 X^\m \del_0 X^\n + \del_1 X^\m \del_1 X^\n) =
{1\ov 4} G_{\m\n} (\del_+ X^\m \del_+ X^\n + \del_- X^\m \del_- X^\n)\ .
\ee
We will assume an underlying group theoretical structure
with currents $I_\pm = I_\pm^a t_a$ valued in some Lie algebra of a semisimple group $G$.
The generators of the Lie algebra $t^a$, $a=1,2,\dots , \dim(G)$ obey the
commutation rules and normalization
\be
[t^a,t^b]= i f^{ab}{}_c t^c\ ,\qq {\rm Tr}(t^a t^b)= \d^{ab}\ .
\label{tatb}
\ee
We will assume that the currents obey
\be
(1+\r) \del_+ I_- + (1-\r) \del_- I_+ =0\ ,
\ee
where $\r$ is a real parameter, and the flat connection identity
\be
\del_+ I_- - \del_- I_+ +  [I_+,I_-]=0\ .
\ee
From these equations we obtain that
\be
\del_+ I_- = -{1-\r\ov 2} [I_+,I_-]\ ,\qq \del_- I_+ = {1+\r\ov 2} [I_+,I_-]\ .
\label{delpm}
\ee
These conditions imply classical integrability.

\subsection{Algebraic and canonical structure}

Our objective is to relate \eqn{delpm} to the equations of motion arising from the $\s$-model \eqn{eq1}.
Following \cite{Balog:1993es} we first introduce the equal time Poisson brackets for the $I^a_\pm$'s
\ba
&& \{ I^a_\pm , I^b_\pm\} = e^2f^{abc}(I_\mp^c -(1+2 x) I_\pm^c)\d(\s-\s') \pm 2e^2 \d^{ab} \d'(\s-\s')\ ,
\nonumber\\
&& \{ I^a_+ , I^b_-\} =-e^2 f^{abc}( I_-^c +  I_+^c)\d(\s-\s')\ ,
\label{iiialf}
\ea
where $e^2$ and $x$ are real parameters.
This algebra obeys the Jacobi identities.
In addition, we assume that the $\s$-model Lagrangian corresponds to the Hamiltonian
\be
H = {1\ov 4 e^2} \int d\s ( I^a_+ I^a_+ + I^a_- I^a_-)\
\label{hamml}
\ee
and that the corresponding evolution
\be
\del_\tau I_\pm^a = \{I^a_\pm,H\}\ ,
\label{del0hi}
\ee
gives rise to the equations for $I_\pm^a$ in \eqn{delpm} with $\r=0$. Details on the derivation of this algebra,
in fact for the more general case with $\r\neq 0$, are given in Appendix A.
Note that for $x=1$ the algebra \eqn{iiialf}
is that corresponding to the PCMs. Furthermore, for large $x$ after an appropriate rescaling of the generators, the same algebra degenerates
into two commuting chiral and antichiral current algebras. The integrable models we will derive
will interpolate between the $\s$-models corresponding to these two cases.

\no
There is no warranty that the usual equal time basic Poisson brackets (for notational
convenience we drop the time dependence)
\be
\{X^\m(\s),P_\n(\s') \} = \d(\s\!-\!\s')\d^\m{}_\n\ ,\quad \{X^\m(\s),X^\n(\s') \}= \{P_\m(\s),P_\n(\s') \} = 0\ ,
\label{xopxp}
\ee
will give rise to \eqn{iiialf}. In fact, demanding that this is the case, would put severe restrictions on the admissible backgrounds.
To further investigate we need to identify the currents $I^a_\pm$ in terms of the $\s$-model data.

\no
We introduce a frame $e^a_\m$ and define chiral and antichiral worldsheet forms as
\be
e^a_\pm = e^a_\m \del_\pm X^\m \ .
\label{eapm}
\ee
Using this frame we construct a spin connection $\om^{ab}$ obeying
\be
de^a + \om^{ab}\wedge e^b = 0\ .
\ee
We will also use the notation $\om^{ab|c}$ defined by writing
\be
\om^{ab} = \om^{ab|c} e^c\ ,
\label{omab}
\ee
as well as the definitions
\ba
&& \om^{ab|c}_\pm = \om^{ab|c}\pm\ha H^{abc}\ ,
\nonumber\\
&& \Om^{abc} = \om^{ab|c} + \om^{ca|b} + \om^{bc|a}\ ,\qq \Om_\pm^{abc}=\Om^{abc} \pm\ha H^{abc}\ ,
\label{ooamb}
\ea
where $H_{abc}$ are the frame components of the antisymmetric field strength three-form
\be
H = d B = {1 \ov 6} H_{abc} e^a \wedge e^b \wedge e^c \ .
\label{hdbb}
\ee
Note also that all the $\Omega$'s are antisymmetric in the three indices.

\no
To compute the equal time Poisson brackets for $e^a_\pm$ we express them as
\be
e_\pm^a = e^{a\m}\left(2 P_\m \pm \ha Q^\pm_{\m\n} \del_\s X^\n\right)\ ,\qq Q^+_{\m\n} = Q_{\m\n}\ ,\quad  Q^-_{\m\n} = Q_{\n\m}\ .
\ee
Then from the Poisson brackets \eqn{xopxp} we obtain after some lengthly computation that
\ba
&& \{e^a_\pm(\s),e^b_\pm(\s')\} = 2 \left(\Om_\mp^{abc}e_\pm^c -\om_\mp^{ab|c}e_\mp^c\right)\d(\s\!-\!\s') \pm 2 \d^{ab}\d'(\s\!-\!\s')\ ,
\nonumber\\
&& \{e^a_+(\s),e^b_-(\s')\} = 2 \left(\om_-^{ca|b}e_+^c -\om_+^{cb|a}e_-^c\right)\d(\s\!-\!\s')\ ,
\label{epembra}
\ea
an algebra which has appeared before in \cite{deBoer:1996eg}.
The next step is to relate $e^a_\pm$ to the currents $I^a_\pm$.
Simply identifying them turns out to be inconsistent. To make progress we introduce another frame $\tilde e^a$
and the analog of \eqn{eapm}, i.e. $\tilde e^a_\pm = \tilde e^a_\m \del_\pm X^\m$. Then
we associate $I_+^a$ with the chiral and $I_-^a$ with the antichiral worldsheet forms as
\be
 I_+^a = 2c_2 \tilde e^a_+\ ,\qq I_-^a =-2 c_2 e^a_- \ ,
\label{imbhfr}
\ee
where the overall constants and sign, as we shall see later, are chosen so that \eqn{epembra} gives rise to \eqn{iiialf}.

\no
The two frames should be  related by a Lorentz transformation $\L$ as
\be
\tilde e^a = \L^{ab} e^b \ ,\qq \L\L^T = \L^T \L =\mathbb{1}\ ,
\label{ela}
\ee
so that they give rise to the same metric.
It turns out that the condition
\be
de^a - \ha (c_1\mathbb{1} + c_2 \L^T)^{ab} f_{bcd} e^c \wedge e^d = 0 \ ,
\label{ea}
\ee
should be satisfied by the frame $e^a$. In addition, the equivalent relation
\be
d\tilde e^a + \ha (c_1 \mathbb{1} + c_2 \L)^{ab} f_{bcd} \tilde e^c \wedge \tilde e^d = 0 \ ,
\label{tea}
\ee
should be satisfied by the frame $\tilde e^a$.

\no
In addition, the antisymmetric field strength three-form in \eqn{hdbb} should assume either one of the two equivalent forms
\ba
H  & = &   -{c_1\ov 6} f_{abc} e^a \wedge e^b \wedge e^c -{c_2\ov 2} f_{abc} \tilde e^a \wedge e^b\wedge e^c
\nonumber\\
& = & -{c_1\ov 6} f_{abc} \tilde e^a \wedge \tilde e^b \wedge \tilde e^c - {c_2\ov 2} f_{abc}  e^a \wedge \tilde e^b\wedge \tilde e^c\ .
\label{Hdb2}
\ea
Then, the components $H$ in the $e^a$ and $\tilde e^a$ bases are given by
\ba
H^{abc} & = &   - c_1 f_{abc} - c_2 (\L_{da} f_{dbc} + \L_{db} f_{dca} + \L_{dc} f_{dab})\ ,
\nonumber\\
\tilde H^{abc} & = & - c_1 f_{abc} - c_2 (\L_{ad} f_{dbc} + \L_{bd} f_{dca} + \L_{cd} f_{dab})\ .
\label{hgrh0}
\ea
Note that, consistency of the above requires the validity of the remarkable identity
\be
H_{abc}  =  \tilde H_{edf} \L^{ea} \L^{db} \L^{fc} \ .
\label{habc}
\ee
This is a generalization of the identity \eqn{dDf} below which is valid for group manifolds.
The conditions \eqn{ea}-\eqn{Hdb2} were essentially found in \cite{Balog:1993es}.

\no
From \eqn{ea} and \eqn{tea} we read off the spin connections and then from \eqn{omab} and \eqn{ooamb} we may compute all the entries
entering in the Poisson algebra \eqn{epembra}. We have collected all of them in Appendix B.
Using these results we can show that the postulated Poisson brackets \eqn{iiialf} arise from the $\s$-model Poisson brackets \eqn{epembra}
after using the identification \eqn{imbhfr}.
In this way we also relate the constants $c_1$ and $c_2$ to the parameters $x$ and $e$.
We find that
\be
c_1 = \left(x+\ha\right)e \ ,\qq c_2 = {e\ov 2}\  .
\label{jhgg}
\ee
Finding solutions to the above conditions is equivalent to specifying the two frames $e^a$ and $\tilde e^a$ from
which the Lorentz matrix $\L$ follows. So far the only solution known is that for the $SU(2)$ case, obtained in
\cite{Balog:1993es} and further studied in \cite{Evans:1994hi}. This solution was obtained essentially after a brute force
computation which clearly cannot be generalized for larger groups.

\subsection{Solving the integrability conditions}

Next we present a general solution to the integrability conditions. The verification of the solution will
be given in full detail in Appendix C.

\no
Consider a Lie algebra for a semisimple group $G$ with generators $t_a$
satisfying \eqn{tatb} and a group element $g\in G$ parametrized by  $\dim(G)$ coordinates.
Then, the left and right invariant Maurer--Cartan forms $L^a$ and $R^a$ are defined as
\be
L^a = -i {\rm Tr}(g^{-1} d g t^a) = L^a_\m dX^\m \ ,\qq R^a = -i {\rm Tr}(d g g^{-1} t^a) = R^a_\m dX^\m \ .
\ee
They obey
\be
dL^a = \ha f^a{}_{bc} L^b \wedge L^c \ ,\qq dR^a =- \ha f^a{}_{bc} R^b \wedge R^c\
\ee
and are related as
\be
R^a = D^{ab} L ^b  \ ,
\ee
where we have defined the matrix $D$ with elements
\be
D_{ab} = {\rm Tr}(t_a g t_b g^{-1}) \ .
\label{Dab}
\ee
This is an orthogonal matrix i.e. $DD^T = D^T D = \mathbb{1}$ and obeys the useful identities
\be
f_{abc}  = f_{def} D_{da}D_{eb} D_{fc} = D_{ad} D_{be} D_{cf} f_{def} \
\label{dDf}
\ee
and
\be
dD_{ab} = D_{ac}f_{cbe}L^e\ .
\ee

\no
We will consider next a $\s$-model of the form \eqn{eq1}. We claim that the solution
to the integrability condition has a metric which can be constructed by using either of the frames\footnote{Note further that, under
inversion of the group element
\be
g\to g^{-1}:\qq L^a \leftrightarrow - R^a\ ,\qq D\to D^T\ ,\qq e^a \leftrightarrow -\tilde e^a \ .
\label{symmmmm}
\ee
Then combining with the worldsheet parity transformation
$\s\to -\s$ the identification \eqn{imbhfr} is consistent with the symmetry \eqn{symmpm}.
}
\be
e^a =  (D - \l \mathbb{1})^{-1}_{ab} R^b \ ,\qq \tilde e^a = (D^T - \l \mathbb{1})^{-1}_{ab} L^b\ .
\label{edtdb}
\ee
The matrix relating the two frames as in \eqn{ela} is
\be
\L = {D-\l \mathbb{1}\ov D^T -\l \mathbb{1}} D^T  = {D-\l \mathbb{1}\ov \mathbb{1} - \l D}\
\label{ljn2}
\ee
and is clearly orthogonal.
In addition, for the antisymmetric tensor we claim that the expression solving the integrability conditions is
\ba
B & = & {1\ov 1-\l^2}\left( B_0 +  \l R^a \wedge e^a\right)
\nonumber\\
 & = & {1\ov 1-\l^2}\left(  B_0 -  \l L^a \wedge \tilde e^a\right)\ ,
\label{dgfbb}
\ea
where $B_0$ is an antisymmetric tensor such that
\be
H_0 = - {1\ov 6} f_{abc} L^a\wedge L^b \wedge L^c = - {1\ov 6} f_{abc} R^a\wedge R^b \wedge R^c\ .
\ee
We have to show that $e^a$, $\tilde e^a$, $\L^{ab}$ and $B$ as given above satisfy the integrability conditions
presented above. This is done in full detail in Appendix C.

\no
Finally the constants $c_1$ and $c_2$ are determined in terms of the parameter $\l$ as
 \be
 c_1 = {1+\l + \l^2\ov 1+\l} \ ,\qq c_2 = {\l\ov 1+\l}\ .
 \label{c1c2}
\ee
Then from \eqn{jhgg} the parameters in the algebra become
\be
e =  {2\l\ov 1+\l}\ ,\qq x  = {1+\l^2\ov 2 \l}\ .
\label{kjk90}
\ee
We will finish this section with a comment concerning the orthogonal matrix $\L$ in \eqn{ljn2}.
We might be tempted to identify this with an expression such as \eqn{Dab} for some
group element $g_0$ parametrized by coordinates $X_0^\m$. However, this is not possible
for all groups since it implies several mutually inconsistent, in general, relations between the $X^\m$'s
parametrizing the group element $g\in G$ and the $X_0^\m$'s. Nevertheless, this is possible in the $SU(2)$ case
and implicitly this was the key factor that the brute force computation of \cite{Balog:1993es} succeeded.
As a related comment note that in the $SU(2)$ case in the identity \eqn{habc} the terms with coefficients $c_1$ and $c_2$
in the components $H_{abc}$ and $\tilde H_{abc}$ in \eqn{hgrh0}
transform to each other separately, whereas for general groups they mix.

\section{Origin of integrability}

We would next derive the integrable $\s$-models we have presented,
that is \eqn{edtdb} and \eqn{dgfbb}. This is indeed necessary since these expressions
are quite complicated and definitely they were not guessed.

\no
Our guide will be the exact CFT coset models $G/H$ \cite{coset} which have
a Lagrangian formulation in terms of gauged WZW models \cite{gwzwac}.
In these constructions one starts with a WZW model action having a $G_L\times G_R$ current algebra symmetry
and gauges a diagonal subgroup $H$. This requires the introduction of non-dynamical gauge fields in the corresponding Lie algebra.
After integrating these gauge fields out we obtain a $\s$-model corresponding to the $G/H$ CFT. This
has a reduced group of isometries as compared to the original WZW action. In fact if $H$ is the maximal subgroup of $G$ then
the resulting space has no isometries at all. Nevertheless, these symmetries can be realized at a string theoretical level.

\no
Inspired by the above we recall the PCM action \cite{Polyakov:1975rr}
\be
S_{\rm PCM}(\tilde g) =-{\kappa^2\ov \pi} \int {\rm Tr}(\tilde g^{-1} \del_+ \tilde g \tilde g^{-1} \del_- \tilde g)\ ,\qq \tilde g\in G\ ,
\label{action1n}
\ee
where $\k^2$ is an overall coupling constant. This model has a global $G_L\times G_R$ symmetry and is integrable.
Recall also the WZW action
\be
S_{\rm WZW}(g) = -{k\ov 2\pi} \int {\rm Tr}(g^{-1} \del_+ g g^{-1} \del_- g) + {i k\ov 6\pi} \int_B {\rm Tr}(g^{-1}dg)^3\ ,\qq g\in G\ ,
\label{dhwzw}
\ee
which has a $G_{\rm L,cur}\times G_{\rm R,cur}$ current algebra symmetry with the overall positive integer $k$ being
the central extension of the current algebra.
By being an exact CFT this model is obviously integrable.

\no
We conclude that the sum of the two actions \eqn{action1n} and \eqn{dhwzw} corresponds to an integrable model with a
$2 \dim(G)$ target space.

\subsection{Gauging the symmetry}

We will gauge the $G_{\rm L}\times G_{\rm diag, cur}$ subgroup of the above symmetry where $G_{\rm diag, cur}$ is the diagonal
subgroup of  $G_{\rm L,cur}\times G_{\rm R,cur}$. The resulting model will still have a global symmetry group
corresponding to $G_{\rm R}$. To proceed with the gauging we first replace in the $S_{\rm PCM}$ action derivatives
by covariant derivatives as
\be
\del_\pm  \tilde g \to D_\pm \tilde g = \del_\pm \tilde g - A_\pm \tilde g \
\label{reldecov}
\ee
and denote the corresponding action by $S_{\rm gPCM}(g;A)$. For the WZW action this minimal substitution does not work. Nevertheless,
the answer is provided by the gauged WZW action for $g\in G$ at level $k$
\ba
&& S_{\rm gWZW}(g;A)= k S_{\rm WZW}(g) + {k\ov \pi}
\int {\rm Tr}(i A_- J_+ - i A_+ J_- + A_- g A_+ g^{-1} - A_-A_+ )
\nonumber\\
 && \phantom{xxxxxxxxx} = k S_{\rm WZW}(g) + {k\ov \pi} \int i A_-^a J_+^a - i A_+ J_-^a +A_+^a (D^T-\mathbb{1})_{ab}A_-^b\ ,
\label{gwzw}
\ea
where
\be
J^a_+ = -i {\rm Tr}(t^a \del_+ g g^{-1}) = R^a_\m \del_+ X^\m \ ,\qq J^a_- = -i {\rm Tr}(t^a g^{-1} \del_- g )= L^a_\m \del_- X^\m\ .
\ee

\no
The total action
\be
S_{\rm tot}(\tilde g,g;A)= S_{\rm gPCM}(\tilde g;A) + S_{\rm gWZW}(g;A)\ ,
\label{gxhan}
\ee
is invariant under the transformation
\be
\tilde g\to \L^{-1} \tilde g\ ,\qquad g\to \L^{-1} g \L\ ,\qquad A_\pm\to \L^{-1}A_\pm \L - \L^{-1}\del_\pm \L \ ,
\label{khgk2}
\ee
for a group element $\L(\s^+,\s^-)\in G$.

\no
Due to the gauge symmetry we should fix $\dim(G)$ parameters among those in the group elements $\tilde g, g\in G$.
Since the group $G$ acts freely on $\tilde g$, we choose the gauge
\be
\tilde g=\mathbb{1}\ .
\label{gauf}
\ee
Since the group $G$ acts with no fixed points we expect that the resulting background will have no singularity.
The total action becomes
\be
S_{\rm g.f.}(g;A) = S_{\rm WZW}(g) +  {1\ov \pi} \int {\rm Tr}(- i k  A_+  J_- + i k J_+ A_- -A_+ M A_-)\ .
\label{gaufix}
\ee
where we have defined the matrix
\be
 M =\k^2 \mathbb{1} - k(D^T-\mathbb{1}) = (k+\k^2) (\mathbb{1}-\l D^T)\ ,
 \label{dskf3jh}
\ee
with the parameter
\be
\l = { k\ov k + \k^2 }\ ,\qq 0<\l<1\ .
\label{dkdjfh11}
\ee
This will be identified with the parameter appearing in the solutions \eqn{edtdb} and \eqn{dgfbb} for the integrability conditions.
It is not hard to show that the gauge fixed action \eqn{gaufix} has a global symmetry given by
\be
g\to \L_0^{-1} g \L_0 \ ,\qq A_\pm \to  \L_0^{-1} A_\pm \L_0 \ ,\qq \L_0 \in G\ .
\label{jhg19}
\ee

\no
Integrating out the gauge fields gives
\be
A_+^a =  i k M^{-1}_{ba} J^b_+ \ , \qq A_-^a = -i k M^{-1}_{ab} J^b_-  \ .
\ee
Substituting back into the action we get the Lagrangian
\be
S(g) = S_{\rm WZW}(g) + {1\ov \pi} {k^2\ov \k^2+ k } \int J_+^a (\mathbb{1}-\l D^T)^{-1}_{ab}J_-^b  \ .
\label{tdulalmorev2}
\ee
This action is invariant under the global symmetry for $g$ in \eqn{jhg19}. In addition, clearly the
corresponding background is non-singular. \footnote{For $\l \neq 1$ the inverse of the matrix $D - \l \mathbb{1}$
exists for all the $X^\m$'s parametrizing $g\in G$.
The reason is that $D$ is an orthogonal real matrix and therefore all of its eigenvalues lie in the unit circle.
}

\subsubsection{Extracting the $\s$-model fields}

From the above action \eqn{tdulalmorev2} we extract the metric as
\ba
 ds^2 & = & {k\ov 2\pi} L^T L + {k^2\ov 2\pi} L^T(D^T M^{-1} + M^{-T} D) L
\nonumber\\
& = &  {k\ov 2\pi} L^T\left[ \mathbb{1} + \l (D-\l \mathbb{1})^{-1} +\l (D^T-\l \mathbb{1})^{-1}\right]L
\\
 & = & {k\ov 2\pi} (1-\l^2) e^a e^a = {k\ov 2\pi} (1-\l^2) \tilde e^a \tilde e^a\ ,
\nonumber
\ea
where $e^a$ and $\tilde e^a$ are the frames given in \eqn{edtdb}. We have also used the notation $M^{-T}$
for $(M^{-1})^T$ and have
disregarded the overall scale $\displaystyle  {k\ov \pi} (1-\l^2)$. Hence we should respect this in
extracting the antisymmetric tensor as well.

\no
To compute the 3-form field strength we write the antisymmetric tensor as
\be
B={1\ov 1-\l^2} \left( B_0 +{\l\ov 2} L^T\left[ (D-\l \mathbb{1})^{-1} - (D^T-\l \mathbb{1})^{-1}\right]\wedge L\right) \ ,
\ee
where $B_0$ is the antisymmetric tensor corresponding to the WZW model.
After, some manipulations this expression takes the form given in \eqn{dgfbb}.

\no
In conclusion, our models originate by gauging the combined actions for the PCM and WZW models for a semisimple group $G$.
This explains in a natural way their integrability properties.

\subsection{The non-Abelian T-dual and CFT limits}

It has been noted in \cite{Sfetsos:1994vz} and further elaborated in \cite{Polychronakos:2010hd} that in the limit in
which the group element approaches the identity and at the same time the level $k$ becomes extremely large in a correlated way,
the action \eqn{tdulalmorev2} becomes that for non-Abelian T-duality. In our case this should be the non-Abelian T-dual
of the PCM action \eqn{action1n} with respect to the $G_{\rm L}$ group. Indeed, let
\be
g = \mathbb{1} + i {v\ov k} + {\cal O}\left( 1\ov k^2 \right)\ , \qq v = v_a t^a\ .
\label{nonbba}
\ee
In that limit we have that
\be
J_\pm^a = {\del_\pm v\ov k} +  \dots  \ ,\qq
D_{ab} = \d_{ab} + {f_{ab} \ov k} +  \dots \ ,
\ee
where the dots stand for subleading terms in the $\displaystyle {1\ov k}$ expansion and
\be
f_{ab} = f_{ab}{}^c v_c\ .
\label{fab}
\ee
Then the action \eqn{gwzw} becomes
\be
S_{\rm gWZW} = -{i\ov \pi} \int {\rm Tr}(v F_{+-}) +  {\cal O}\left( 1\ov k \right) \  ,
\quad F_{+-} = \del_+ A_- - \del_- A_+ -[A_+,A_-]   \ .
\ee
Note that the contribution of the WZW action $S_{\rm WZW}$ in this limit is subleading.
Hence in the $k\to \infty$ limit, the action \eqn{gxhan} becomes the starting point for the usual non-Abelian T-duality action.
In accordance, the action \eqn{tdulalmorev2} becomes that resulting from the non-Abelian T-dual transformation applied to the
PCM \eqn{action1n}, i.e.
\be
S_{\rm non\!-\!Abel}(v) = {1\ov \pi} \int \del_+ v^a (\mathbb{1} + f)^{-1}_{ab}\del_- v^b  +  {\cal O}\left({1\ov k}\right)\ .
\label{ndfka}
\ee
In the opposite limit when $k\ll\k^2 $ we have that
\be
S  = S_{\rm WZW} + {k^2\ov \pi \k^2} \int J^a_+J_-^a + {\cal O}(k^3)\ .
\label{jfkgbibi}
\ee
revealing the behavior of a perturbed WZW model by a current bilinear.\footnote{This perturbation is not exactly marginal.
since it violates the criteria of \cite{Kadanoff:1978pv,Chaudhuri:1988qb}.}
For the $SU(2)$ case this was noted in \cite{Evans:1994hi}.

\no
We emphasize that the proof that our models are integrable includes the non-Abelian T-dual limit action \eqn{ndfka} as well.
In view of its particular importance, we present in
Appendix D an independent proof solely for the case of the non-Abelian T-duality.

\section{The $SU(2)$ example}

Consider the case in which the group $G=SU(2)$ and the corresponding $\s$-model is the round
$S^3$.
From our general construction we know that there will be an $SU(2)$ isometry in the resulting
background. It will be convenient to parametrize the group element $g\in SU(2)$ in such a way that this symmetry is manifest
in the background. The appropriate parametrization is given by
\be
g=e^{i \a_i \s_i}=
\pmatrix{\cos{\a} + i \sin\a \cos\b & \sin\a \sin\b \ e^{-i\g} \cr
-\sin\a \sin\b \ e^{i\g} & \cos{\a} - i \sin\a \cos\b \cr}\ ,
\label{su2g2}
\ee
where we have defined
\be
\a_1 = -\a \sin\b \sin\g \ ,\qq  \a_2 = \a \sin\b \cos\g\ ,\qq   \a_3 = \a \cos\b\ .
\ee
Then
\ba
&& S_{\rm WZW} = {k\ov \pi} \int \del_+ \a\del_- \a +\sin^2\a (\del_+ \b \del_- \b + \sin^2\b \del_+ \g\del_-\g)
\nonumber\\
&& \phantom{xxxxxxx}
-(\a-\sin\a\cos\a)\sin\b (\del_+\b \del_- \g - \del_+\g \del_- \b) \ .
\label{jhk2}
\ea
Working out the details of \eqn{tdulalmorev2} we obtain a $\s$-model with metric and antisymmetric tensor given by
\ba
&& ds^2 =  {k(\k^2+2k)\ov \k^2}
\left( d\a^2 + \k^4 {\sin^2\a\ov \D} ds^2(S^2)\right)\ , \qq \D = \k^4 +4k(\k^2+k)\sin^2\a \ ,
\nonumber\\
&&
B = -k \left(\a - \k^4 {\sin \a \cos\a\ov \D}\right) {\rm Vol}(S^2)\ .
\label{dsbdn}
\ea
where $S^2$ is the unit two-sphere with metric $\displaystyle ds^2(S^2) = d\b^2 + \sin^2\b d\g^2$.
This background is non-singular in accordance with the general comment we made below \eqn{gauf} and footnote 3.

\no
According to our general discussion, for $k\ll \k^2$ the fields in the background \eqn{dsbdn} become indeed those
for the $SU(2)$ WZW model in \eqn{jhk2}.
The limit resulting into the non-Abelian T-duality background is just
\be
\a = {r\ov 2k}\ ,\qq {\rm then}\ k\to \infty\ ,
\ee
since then the group element \eqn{su2g2} can be expanded around the identity.
This limiting procedure gives
\be
ds^2 =\ha \left(dr^2 + {r^2\ov r^2+1} ds^2(S^2)\right) \ ,\qq
B = -\ha {r^3\ov r^2+1} {\rm Vol}(S^2)\ .
\ee
This is the non-Abelian T-dual of the $SU(2)$ PCM which in fact has been embedded in supergravity.
It was shown in \cite{Sfetsos:2010uq} that when supported by appropriate flux fields it is a solution of massive IIA-supergravity
and that it represent the non-Abelian T-dual of the background corresponding to the near horizon of the D1-D5 brane system.

\no
For completeness in Appendix E we establish a precise relation between the background \eqn{dsbdn} and that in
\cite{Balog:1993es}.

\section{General models}

The above construction can be generalized to $\s$-models with global symmetry $G_{\rm L}$. These
have the general form
\be
S(g,X) ={1\ov \pi} \int Q_{\a\b} \del_+ Y^\a\del_-Y^\b +  Q_{\a a} \del_+ Y^\a L_-^a + Q_{a\a} L_+^a  \del_- Y^\a + E_{ab} L_+^a L_-^b \ ,
\label{action1}
\ee
where the fields $Y^\a$, $\a=1,2,\dots d$ do not transform under the global symmetry. The
coupling matrices $Q_{\a\b}$, $Q_{\a a}$,  $Q_{a\a}$ and $E_{ab}$ may depend on the $Y^\a$'s.

\no
As before we add to this the WZW action \eqn{dhwzw} and then we gauge the $G_{\rm L}\times G_{\rm diag,cur}$ subgroup on the total
action.
For \eqn{action1} we just replace the derivatives by covariant derivatives as in \eqn{reldecov}.
To this we add the gauged WZW action \eqn{gwzw}. Then, the total action is invariant under \eqn{khgk2} and the spectator
fields $Y^\a$ are left invariant. After choosing the gauge fixing condition \eqn{gauf} the total action becomes
\ba
&&S_{\rm g.f.} = S_{\rm WZW}(g)+ {1\ov\pi} \int Q_{\a\b} \del_+ Y^\a \del_- Y^\b  - A_+^a M_{ab} A_-^b
\nonumber\\
&& \phantom{xxxxx} +  i A_+^a(Q_{a\a}\del_- Y^\a - k J_-^a) + i (Q_{\a a}\del_+ Y^\a + k J_+^a)A_-^A\ ,
\ea
where we have defined the matrix
\be
 M =E - k(D^T-\mathbb{1})\ ,
 \ee
which, in this general case, replaces the definition \eqn{dskf3jh}.

\no
Integrating out the gauge fields gives
\be
A_+^a =  i M^{-1}_{ba} (k J^b_+ + Q_{\a b}\del_+ Y^\a ) \ ,
\quad A_-^a = -i M^{-1}_{ab} (k J^b_-  - Q_{b\a} \del_- Y^\a)\ .
\ee
Substituting back into the action we get the dual Lagrangian
\be
S= S_{\rm WZW}(g) + {1\ov \pi} \int Q_{\a\b}\del_+ Y^\a \del_- Y^\b  +  (k J_+^a  + \del_+ Y^\a Q_{\a a})M^{-1}_{ab}
(k J_-^b  - Q_{b\b} \del_- Y^\b )\ .
\label{tdulalmore}
\ee
Whether or not this $\s$-model action is integrable depends on the details of the various couplings.
In particular, in the absence of spectator fields the form of the matrix $E_{ab}$ is crucial for that.

\no
Note that in the limit \eqn{nonbba} we obtain from \eqn{tdulalmore} the action
\be
S= {1\ov \pi} \int Q_{\a\b}\del_+ Y^\a \del_- Y^\b  +  (k \del_+ v^a  + \del_+ Y^\a Q_{\a a})(E+f)^{-1}_{ab}
(\del_- v^b  - Q_{b\b} \del_- X^\b )\ .
\label{tdulalmnon}
\ee
This is nothing but the action arising from the non-Abelian T-dual of \eqn{action1} with respect of the
$G_{\rm L}$ symmetry. In the opposite limit in which $k\ll \k^2$ we have that
\ba
&& S= S_{\rm WZW}(g) + {1\ov \pi} \int Q_{\a\b}\del_+ Y^\a \del_- Y^\b
\nonumber\\
&& \phantom{xxxx}
+ {1\ov \pi} \int  (k J_+^a  + \del_+ Y^\a Q_{\a a})E^{-1}_{ab} (k J_-^b  - Q_{b\b} \del_- Y^\b ) + {\cal O}(k^3)\ .
\label{tdulalmksm}
\ea

\no
In the absence of spectator fields \eqn{tdulalmore} becomes
\be
S= S_{\rm WZW}(g) + {k^2\ov \pi} \int J_+^a M^{-1}_{ab} J_-^b \ .
\label{tdulalmorenn}
\ee
After setting $E = \k^2 \mathbb{1}$ this action reduces to \eqn{tdulalmorev2}.
We also note that the form of the action \eqn{tdulalmorenn} has appeared before in \cite{Tseytlin:1993hm} in studies of gauged-WZW-like actions.

\no
Finally, we mention that a class of integrable models corresponding to specific
choices for the matrix $E_{ab}$ depending on the group $G$, were constructed in \cite{Klimcik:2008eq}.
It will be interesting to investigate if the model \eqn{tdulalmorenn} is integrable when these matrices are used.

\section{New integrable interpolations with coset models}

We would like to extend the discussion to cases in which in the action \eqn{action1} the group index $a$ runs over a coset space
$G/H$. Since the group action is no longer free that means that one cannot fix all the parameters in the group element $\tilde g$ to specified values,
i.e. the gauge fixing \eqn{gauf} is not possible and one has to also fix some of the parameters in the group element $g$ parametrizing
the WZW action. Here we follow a limiting procedure developed in \cite{Sfetsos:1999zm}
and further explored in the context of non-Abelian T-duality transformation for geometries containing coset spaces in
\cite{Lozano:2011kb}. To keep the presentation simple consider the case with no spectator fields.
Let's choose
\be
E = \diag(E_0 , s^2 \mathbb{1}_{\dim{H}})\ ,
\ee
with $E_0$ a matrix in the coset which is $G$-invariant and $s$ is a parameter.
As long as the parameter $s$ is non-vanishing, the resulting action \eqn{tdulalmorenn} is valid since the action of the group $G$ is free.
In the limit $s\to 0$ all quantities with an index in the subgroup $H$ drop out in the original action \eqn{action1} with
no spectator fields.
The effect of this limit is that the action \eqn{tdulalmorenn} develops a gauge invariance under $H$ which allows
to reduce the configuration space from $\dim(G)$ to $\dim(G/H)$.

\no
For small $k$ the $\s$-model action corresponds to the exact CFT gauged WZW model for a group $G$ and a subgroup $H$.
In the limit of large $k$ the result is
the non-Abelian T-dual of the $\s$-model for the coset $G/H$ with respect to $G$.
Based also on the example below,
we expect that the perturbation driving these models away from the conformal point,
will be based on bilinears, analogous to \eqn{jfkgbibi}, of the classical parafermions \cite{Bardacki:1990wj,Bardakci:1990ad},
which are classical counterparts of the quantum parafermions constructed earlier in \cite{Fateev:1985mm}
(for the $SU(2)_k/U(1)_k$ coset case).

\subsection{Integrable models based on the $SU(2)/U(1)$ exact CFT}

Consider again the case with $G=SU(2)$ and take the matrix
\be
E = \diag(1,1,s^2)\ ,
\ee
where the spit of the indices is such that the third entry corresponds to the $U(1)$ subgroup of $SU(2)$,
having in mind to take at the end of the computation the limit $s\to 0$.
In this case it is more convenient to parametrize the group element as
\be
g = e^{i (\phi_1 -\phi_2) \s_3/2} e^{i \om \s_2}e^{i (\phi_1 +\phi_2) \s_3/2}\ ,
\ee
instead of \eqn{su2g2}.
Then the corresponding WZW action is given by
\ba
&& S_{\rm WZW}(g) = {k\ov \pi}\int \del_+ \om \del_-\om
+ \cos^2\om \del_+\phi_1 \del_-\phi_1 +  \sin^2\om \del_+\phi_2 \del_-\phi_2
\nonumber\\
&& \phantom{xxxxxxx}
-\ha \cos 2\om (\del_+ \phi_1 \del_-\phi_2 - \del_+ \phi_2 \del_-\phi_1)\ .
\label{dgkj91}
\ea
After taking the limit $s\to 0$ the coordinate $\phi_2$ drops out.
The result is the $\s$-model action
\ba
&& S
= {k\ov\pi(2k+1)}\int \Big[ \del_+\om \del_-\om + \cot^2\om \del_+ \phi_1\del_- \phi_1
\nonumber\\
&& \phantom{}
+ 4 k(k+1)
(\cos\phi_1 \del_+ \om + \sin\phi_1 \cot\om \del_+ \phi_1)
(\cos\phi_1 \del_- \om + \sin\phi_1 \cot\om \del_- \phi_1)\Big]\ .
\label{clpp2}
\ea

\subsubsection{Relation to integrable perturbations and non-Abelian T-duality}

The action \eqn{clpp2} has an exact CFT interpretation. For small $k$ the dominant term is
the first line which is nothing but the $\s$-model action
\be
S_{\rm CFT} ={k\ov \pi} \int \del_+\om \del_-\om + \cot^2\om \del_+ \phi_1\del_- \phi_1\ ,
\label{scgy}
\ee
for the exact $SU(2)/U(1)$ coset CFT \cite{Bardacki:1990wj}.
The term in the second line is a bilinear which in fact can be written with the help of the
classical parafermions given by
\begin{equation}
\label{classcp} \psi =\big(\del_+ \om + i \cot\om\
\del_+ \phi_1\big) e^{-i(\phi_1+\tilde \phi_1)}\ ,\qquad  \psi^\dagger
=\big(\del_+ \om - i \cot\om\ \del_+ \phi_1\big)
e^{i(\phi_1+\tilde\phi_1)} \
\end{equation}
and
\begin{equation}
\label{classcp1} \bar \psi = \big( \del_- \om + i
\cot\om\   \del_- \phi_1\big) e^{-i(\phi_1-\tilde\phi_1)}\ ,
\qquad  \bar \psi^\dagger = \big(
\del_- \om - i \cot\om\ \del_- \phi_1\big)
e^{i(\phi_1-\tilde\phi_1)}\ .
\end{equation}
In general the parafermions originate from the currents with coset indices of the $G_k$ theory.
In the gauged theory they are
dressed with gauge fields that render them gauge-invariant. In our case this dressing is done by
the phase $\tilde \phi_1$ which is a
non-local function of the variables $\om$ and $\phi_1$. Its
explicit expression is not needed here (see, for instance,
\cite{Marios Petropoulos:2005wu}), but it is necessary for ensuring on-shell chiral and
anti-chiral conservation of the parafermions
\begin{equation}
\label{dik1}
\del_- \psi = \del_- \psi^\dagger =0\ , \qq
\del_+ \bar\psi = \del_+  \bar\psi^\dagger =0\ .
\end{equation}
Then the action \eqn{clpp2} can be written and then expanded as
\ba
&& S  ={1\ov 2k+1} \left[[1+2k(k+1)] S_{\rm CFT} + {k^2(k+1)\ov \pi} \int (\psi\bar \psi
+ \psi^\dagger\bar \psi^\dagger)\right]
\nonumber\\
&& \phantom{xxxx}
 = S_{\rm CFT} +{k^2\ov \pi} \int (\psi\bar \psi + \psi^\dagger \bar \psi^\dagger) +{\cal O}(k^3)\ .
\ea
This is indeed the exact CFT action \eqn{scgy} perturbed by parafermion bilinears which is a
relevant perturbation since the parafermions have conformal dimension $1-1/k$. Furthermore this perturbation was
studied in \cite{Fateev:1991bv} where it was shown to be integrable, massive and argued that in the
$k\to \infty$ limit the model flows under the renormalization group to the $O(3)$ $\s$-model.
Testing this with our action \eqn{dgkj91}, consider the rescaling followed by the $k\to \infty$ limit
\be
\phi_1 = {x_1\ov 2k}\ ,\qq \om = {x_2\ov 2 k}\ , \qq {\rm and\ then}\ k\to \infty\ .
\ee
We obtain that
\be
ds^2 = {dx_1^2\ov x_2^2}+ \Big(dx_2 + {x_1\ov x_2}dx_1\Big)^2\ ,
\label{dhlk22}
\ee
which is the non-Abelian T-dual of $S^2$ (see, for instance, \cite{Lozano:2011kb}).
This is consistent with the work of \cite{Fateev:1991bv} since
the PCM and its non-Abelian T-dual are supposed to be equivalent. In particular, the $\s$-model corresponding
to \eqn{dhlk22} has the same $S$-matrix as the $O(3)$ model.

\no
Note that if we had dressed the parafermion bilinears so that the perturbation was exactly
marginal, we would have flown to different CFTs as in \cite{Petropoulos:2006py} and in \cite{Marios Petropoulos:2005wu,Fotopoulos:2007rm}.

\section{Concluding remarks}

We have constructed a large class of new integrable theories. They have an underlying group theoretical structure
and are obtained by gauging a subgroup of the combined actions for the WZW model and the PCMs.
Our theories interpolate between exact CFT WZW models and the non-Abelian T-duals of PCM.
We generalized this construction by replacing the PCM part of the $\s$-model action by a more general
one having less symmetry and in addition spectator fields.
Using this action we constructed models interpolating between exact
coset CFTs realized by gauged WZW models and non-Abelian T-duals of geometric cosets.
We presented evidence that these models are integrable as well.

\no
It is interesting to investigate the behavior of our models under the renormalization group flow.
The work of \cite{Balog:1993es,Evans:1994hi} for the simple $SU(2)$ case suggests that
they should flow from the CFT point to the
non-Abelian T-duality point as we go from the UV to the IR.
This is in accordance with the fact that the perturbation from the CFT side is driven by relevant operators.
It will be interesting to demonstrate this explicitly and in general.

\no
It is important to construct models with non-vanishing parameter $\r$. These are allowed by the algebra
\eqn{iiialfgen}, with the parameters given by \eqn{allcon}, but no example is known.
As a related comment, note that the anisotropic extension of the PCM for $SU(2)$, i.e. \eqn{action1} with no spectators, $G=SU(2)$,
and $E_{ab}$ diagonal, is also integrable \cite{Cherednik:1981df}. Our construction suggests that the \eqn{tdulalmorenn}
is a promising candidate for an integrable model that would interpolate between the anisotropic PCM model and the WZW model for $SU(2)$.
In addition, we also mention that some general conditions for a $\s$-model to be integrable were discussed in
\cite{Mohammedi:2008vd}. It would be interesting to see if models of the type discussed
in section 6 could provide solutions to these conditions.

\no
An original motivation for this work relates to our quest for a better understanding of global issues in backgrounds obtained by
non-Abelian T-duality transformations.
A major unresolved such issue concerns the range of the variables in the T-dual $\s$-model backgrounds.
According to the proposal of \cite{Sfetsos:1994vz} such models can be thought of as effectively describing
high spin sectors of some parent theories \cite{Polychronakos:2010hd}.
In this paper we enriched this idea by showing that non-Abelian T-duals can be at the end point of a whole line
of integrable deformations of exact CFTs. In a generic point in this deformation all variables are compact.
As the deformation parameter becomes infinitely large one can effectively rescale the compact variables and zoom into the manifold.
From this point of view
the non-compactness of the variables in the $\s$-model corresponding to the non-Abelian T-dual theory is explained.

\no
It will be very interesting to investigate a possible embedding of our construction to type-II supergravity with
non-trivial Ramond--Ramond fields. At the limit of non-Abelian T-duality this is possible \cite{Sfetsos:2010uq,Lozano:2011kb}
and there has been related studies and applications in the context of the gauge/gravity
correspondence \cite{Itsios:2012zv,Lozano:2012au,Itsios:2013wd,Barranco:2013fza,Lozano:2013oma}.
If such an embedding can be done physical questions related for instance to the fate of charges of D- or p-branes after
the non-Abelian T-duality transformation can be addressed with confidence.
This is due to the resolution of the global issues of the $\s$-model variables as described above which will render
certain integrals that appear in this type of computations finite with no need for a regularization.
A related comment concerns the possibility of constructing integrable deformations of superstring actions of direct
interest in the AdS/CFT correspondence based on our models.
Such a construction was done recently for a different integrable deformation of the $AdS_5\times S^5$ superstring
in \cite{Delduc:2013qra} for which an $S$-matrix has also been proposed before in a more general context in \cite{Hoare:2011wr}
and further checked in perturbation theory in \cite{Arutyunov:2013ega}.

\no
Finally, it is important to shed more light into the underlying algebraic structure of the integrable models we have found.
The Poisson brackets in \eqn{iiialf} were fixed by making an ansatz and then imposing
for consistency the Jacobi identities and that they should encode the equations of motion.
It should be possible to derive \eqn{iiialf} from the Poisson brackets for the WZW models and the PCMs by imposing
appropriate constraints arising from the gauging procedure.
This should be very helpful in understanding and exploring further the integrability structure of our theories.
Recall also that there is a Yangian algebra \cite{Drinfeld:1985rx} (for reviews see \cite{Bernard:1992ya,Torrielli:2010kq})
that the classically conserved non-local charges of the PCM obey \cite{MacKay:1992he}.
Quantum effects to their conservation properties have also been explored (for various aspects of the algebras of
charges in PCM see \cite{Abdalla:1986xb,Evans:1999mj}).
For the case of our models it would be very interesting to show that they have hidden symmetries encoded in Yangian algebras
since a preservation of the Yangian symmetries after a deformation is not immediate.
In that respect we note that in a somewhat similar, albeit simpler case, according to the work of
\cite{Kawaguchi:2010jg} the Yangian symmetries are preserved for the squashed $S^3$ $\s$-model.
The derivation of our integrable models by a
a gauging procedure should be instrumental in making further progress in that direction as well.

\section*{Acknowlegments}
I would like to thank G. Itsios for discussions related to this paper and D.C. Thompson for a critical reading of the manuscript.
This research is implemented under the {\it ARISTEIA} action of the
{\it operational programme education and lifelong learning} and is co-funded by the European Social
Fund (ESF) and National Resources.

\appendix
\section{Derivation of the Poisson brackets}

Consider a $\s$-model with Hamiltonian given by \eqn{hamml}
and assume the ansatz for the equal time Poisson brackets
\ba
&& \{ I^a_\pm , I^b_\pm\} = f^{abc}(a_\pm I_-^c + b_\pm I_+^c)\d(\s-\s') + c_\pm \d^{ab} \d'(\s-\s')\ ,
\nonumber\\
&& \{ I^a_+ , I^b_-\} = f^{abc}(d I_+^c + f I_-^c)\d(\s-\s')\ .
\label{iiialfgen}
\ea
In addition, we assume that the $\s$-model Lagrangian corresponds to the Hamiltonian \eqn{hamml}.
A possible central extension in the second line Poisson bracket in \eqn{iiialf} has not been included since that
would have been incompatible with the evolution \eqn{delpm}.

\no
We will determine the constants $a_\pm,b_\pm,c_\pm,d$ and $f$ by demanding that the
Jacobi identities are obeyed and that the Hamiltonian evolution \eqn{del0hi}
gives rise to the equations for $I_\pm^a$ in \eqn{delpm}.
It is natural to demand that under worldsheet parity we have that
\be
\s\to -\s \quad \Longrightarrow \quad I^a_\pm \to I^a_\mp\ ,
\label{symmpm}
\ee
we have that
\be
a_\pm (-\r) = b_\mp(\r)\ ,\qq c_+(-\r)= - c_-(\r)\ ,\qq d(-\r)= f(\r)\ .
\label{apfkj4}
\ee
The algebra \eqn{iiialf} has to obey the Jacobi identities. Due to the above symmetry the only independent
ones are $\{\{I^a_+,I^b_+\},I^c_+\} + \cdots=0$ and $\{\{I^a_+,I^b_+\},I^c_-\} + \cdots=0$.
The first one is trivially satisfied. The second gives the conditions
\be
a_+ a_- + f b_+ - d a_+ -f^2 =0 \ ,\qq a_+ b_- - fd =0 \ .
\label{appjfg32}
\ee
Recalling the time evolution \eqn{del0hi} we compute
\be
\del_0 I^a_+ = \cdots = {a_+-d\ov 2 e^2} f^{abc} I^b_+ I^c_- + {c_+\ov 2e^2} \del_\s I^a_+\ .
\ee
Using that $[t^a,t^b]=f^{abc}t^c$  we see that the second of \eqn{delpm} is reproduced if the conditions (setting temporarily $e^2=1$)
\be
c_+ = 2\ ,\qq a_+ - d =2 (1+\r)\ ,
\ee
are obeyed. Using the second of them as well as the one that follows by using \eqn{apfkj4},
the last condition in \eqn{appjfg32} can be written as
\be
(1+\r)f + (1-\r) d + 2(1-\r^2)= 0\ .
\ee
Keeping in mind that $f(\r)=d(-\r)$ we easily see that the solution is $f = -1+\r$. Then
\eqn{appjfg32} becomes $(1+\r)a_- - (1-\r)b_+ + 4\r=0$. Given that $b_+(-\r)=a_-(\r)$ we have that
$b_+ = -(1+2 x) + (1-2 x)\r$, where $x$ is an arbitrary real parameter.
Collecting everything, we have that the most general solution is given by
\ba
&& f = -(1-\r)e^2 \ ,\qquad d = -(1+\r)e^2 \ ,
\nonumber\\
&& a_+ = (1+\r)e^2 \ ,\qq b_+ = -\left[(1+2 x) - (1-2 x)\r\right]e^2\ ,\qq c_+ = 2e^2 \ ,
\label{allcon}
\\
&& a_- = -\left[(1+2 x) +(1-2 x)\r\right]e^2 \ ,\qq b_- = (1-\r)e^2\ ,\qq c_-= - 2e^2\ .
\nonumber
\ea
Note that these contain, besides $\r$, the overall scale factor $e^2$ and the free parameter $x$.
For $\r=0$ we obtain the algebra \eqn{iiialf}.

\section{Quantities related to the spin connections}

In this appendix we collect the expressions needed to specify completely the
right hand sides of the Poisson bracket algebras in \eqn{epembra}.

\no
From \eqn{ea} and \eqn{tea} we read off the spin connections and then from \eqn{omab} that
\ba
&& \om^{ab|c} = {c_1\ov 2} f_{abc} - {c_2\ov 2} \left(\L_{dc} f_{dab} - \L_{da} f_{dbc} - \L_{db} f_{dca}\right)\  ,
\nonumber\\
&& \tilde \om^{ab|c} = - {c_1\ov 2} f_{abc} + {c_2\ov 2} \left(\L_{cd} f_{dab} - \L_{ad} f_{dbc} - \L_{bd} f_{dca}\right)\ ,
\ea
from which, using \eqn{ooamb}
\ba
\om_+^{ab|c} = -c_2 \L_{dc}f_{d ab} \ , \qq \om_-^{ab|c} = c_1 f_{abc} + c_2 (\L_{da} f_{dbc} + \L_{db} f_{dca})\ ,
\nonumber\\
\tilde \om_-^{ab|c} = c_2 \L_{cd}f_{d ab} \ , \qq \tilde \om_+^{ab|c} = - c_1 f_{abc} - c_2 (\L_{ad} f_{dbc} + \L_{bd} f_{dca})\ .
\ea
In addition, we find from \eqn{ooamb} that
\ba
&& \Om^{abc} = {3c_1\ov 2} f_{abc} +  {c_2\ov 2} (\L_{da} f_{dbc} + \L_{db} f_{dca} + \L_{dc} f_{dab})\ ,
\nonumber\\
&& \tilde \Om^{abc} = -{3c_1\ov 2} f_{abc} -  {c_2\ov 2} (\L_{ad} f_{dbc} + \L_{bd} f_{dca} + \L_{cd} f_{dab})\
\ea
and
\ba
&& \Om^{abc}_+ = c_1 f_{abc}\ ,\quad \Om^{abc}_- = 2 c_1 f_{abc} + c_2 (\L_{da} f_{dbc} + \L_{db} f_{dca} + \L_{dc} f_{dab})\ ,
\nonumber\\
&& \tilde \Om^{abc}_- = - c_1 f_{abc}\ ,\quad \tilde \Om^{abc}_+ = -2 c_1 f_{abc} - c_2 (\L_{ad} f_{dbc} + \L_{bd} f_{dca} + \L_{cd} f_{dab})\ .
\ea
Finally, the Lorentz transformation $\L$ has to obey
\be
d\L^{ab} = -c_2 f_{adc} \L^{db} e^c + c_2 f_{abc} e^c + c_1 \L^{ae} f_{ebc} e^c + c_2 \L^{ae} f_{edc} \L^{db} e^c\ .
\label{dllit}
\ee

\section{Proving the integrability conditions }

In this appendix we prove that the expressions for the frames \eqn{edtdb} and the antisymmetric
tensor \eqn{dgfbb} solve the conditions \eqn{ea}, \eqn{tea}, \eqn{Hdb2} and \eqn{dllit}.

\no
A simple computation using \eqn{edtdb}  shows that
\be
d\tilde e^a = \ha (D^T-\l \mathbb{1})^{-1}_{ab} f_{bcd} (D^T+\l \mathbb{1})_{ce} (D^T-\l \mathbb{1})_{df} \
\tilde e^e\wedge \tilde e^f \ .
\ee
Using \eqn{dDf} we may further write that
\be
d\tilde e^a = -\ha f^{(2)}_a + {\l(\l-1)\ov 2} (D^T-\l \mathbb{1})^{-1}_{ab}  f^{(2)}_b\ ,
\qq  f^{(2)}_a \equiv f_{abc} \tilde e^b \wedge \tilde e^c\ .
\ee
However, using the identity
\be
(D^T-\l \mathbb{1})^{-1} = {1\ov 1-\l^2} (\L + \l \mathbb{1})\ ,
\ee
we find that $d\tilde e^a$ satisfies \eqn{tea} with the constants being specified as in \eqn{c1c2}.
Similarly we find that $de^a$ satisfies \eqn{ea}.

\no
Next we turn our attention to the antisymmetric tensor. Starting from the second expression in \eqn{dgfbb} we compute that
\ba
&& dB = -{1\ov 6(1-\l^2)} f_{abc} L^a \wedge L^b \wedge L^c - {\l\ov 1-\l^2} dL^a \wedge \tilde e^a + {\l\ov 1-\l^2} L^a \wedge d\tilde e^a
\nonumber\\
&&\phantom{xx}
= -{1\ov 6(1-\l^2)} f_{def} (D^T-\l \mathbb{1})^{da} (D^T-\l \mathbb{1})^{eb} (D^T-\l \mathbb{1})^{fc} \tilde e^a \wedge \tilde e^b \wedge \tilde e^c
\nonumber\\
&&
\phantom{xxxx}
- {\l \ov 2(1-\l^2)} f^{aef} (D^T-\l \mathbb{1})^{eb} (D^T-\l \mathbb{1})^{fc} \tilde e^a \wedge \tilde e^b \wedge \tilde e^c
\\
&& \phantom{xxxx}
-{\l\ov 2(1-\l^2)} f_{bcd} (D^T-\l \mathbb{1})^{ea} (c_1 \mathbb{1} + c_2 \L)^{e d} \tilde e^a \wedge \tilde e^b \wedge \tilde e^c\ .
\nonumber
\ea
After some manipulations
\ba
&& dB = -{1+2\l^3\ov 6(1-\l^2)} f_{abc} \tilde e^a \wedge \tilde e^b \wedge \tilde e^c
\nonumber\\
&& \phantom{xxxx}  + {\l\ov 2(1-\l^2)} f_{abd} \left(\l D - (D-\l \mathbb{1})(c_1\mathbb{1} + c_2 \L)\right)^{cd}
\tilde e^a \wedge \tilde e^b \wedge \tilde e^c\ .
\ea
Using \eqn{ljn2} we find the identity
\be
\l D - (D-\l \mathbb{1})(c_1\mathbb{1} + c_2 \L) = \l^2 \mathbb{1} - (1-\l) \L \ .
\ee
Then we find for $H$ the second line in \eqn{Hdb2}.
A similar computation gives the alternative form given by the first line in \eqn{Hdb2} which is easily seen to be consistent
with \eqn{symmmmm} since when $g\to g^{-1}$, then $H\to -H$.

\section{Integrability and non-Abelian T-duality on PCM}

In this appendix we present an alternative proof of the integrability of the
$\s$-model action \eqn{nonabealdd} corresponding to the PCM action \eqn{action1n}.
The relation between integrability and non-Abelian T-duality has been also examined in
\cite{Mohammedi:2008vd}.

\no
In a two-dimensional Minkowski spacetime consider the action
\be
S = - \int {\rm Tr}\left(\ha j\wedge \star j + v (d j + j\wedge j )\right)\ ,
\label{non-abeldual0}
\ee
where the Hodge dual acts as $ \star dx_0 = dx_1 , \star dx_1 = dx_0 , \star^2 = 1$.

\no
Varying \eqn{non-abeldual0} with respect to $v$ enforces the flatness condition
\be
dj + j\wedge j = 0 \ ,
\label{flatt}
\ee
which is solved as $j=g^{-1}dg$ and putting it
back into the action gives the PCM action
\be
S = -\ha \int {\rm Tr}(j\wedge \star j )\ ,\qq j=g^{-1}dg\ ,
\label{pcmmm}
\ee
equivalent to \eqn{action1n}.
Varying this action with respect to $g$ we get the equation of motion
\be
d(\star j )= 0 \ .
\label{jerw}
\ee
The flatness condition \eqn{flatt} and the equation of motion \eqn{jerw} follow from the
flatness condition for
\be
J(z) = a j + b \star j \ ,\qq a= -{1\ov 4} (z-z^{-1})^2\ ,\quad b= {1\ov 4} (z^2-z^{-2})\ ,
\label{jzolo}
\ee
where $z$ is the spectral parameter.
This one-parameter family of currents leads to an infinite number of concerned charges.

\no
Alternatively we may vary \eqn{non-abeldual0} over $j$ and get the equation
\be
\star j + dv + [j,v] = 0 \ .
\label{eqpcpm}
\ee
This can be solved in light cone components as
\be
j_\pm = \mp {1\ov \mathbb{1}\pm f}\ \del_\pm v \quad \Longrightarrow \quad
j = {f\ov \mathbb{1}-f^2}\ dv - {1\ov \mathbb{1}-f^2}\ \star dv\ ,
\label{jmlpp}
\ee
where $f$ is the matrix with elements \eqn{fab}.
The on-shell action is
\be
S = -\ha \int {\rm Tr}(j\wedge dv) = \int d^2 x\ \del_+ v {1\ov \mathbb{1}-f}\del_- v \ ,
\label{nonabealdd}
\ee
where $d^2 x= dx^+ \wedge dx^-$. This is nothing but the action \eqn{ndfka}.
Varying this action with respect to $v$ we obtain after some algebra the equation
\be
\del_+ j_- - \del_-j_+ + [j_+,j_-] = 0 \quad \Longrightarrow \quad dj + j\wedge j = 0 \ ,
\ee
where $j_\pm$ are given in \eqn{jmlpp}.
Hence the flatness condition \eqn{flatt} is obeyed as an equation of motion.
In addition, from the fact that
\eqn{eqpcpm} can be written as $\star j +\nabla v =0$ with $\nabla(\cdot) = d(\cdot) + [j,.]$
and that $j$ is flat we have that $\nabla (\star j) = -\nabla^2 v = 0 $, which implies that
\be
d(\star j) +[j,\star j] = 0 \quad \Longrightarrow \quad d(\star j)=0\ ,
\ee
since $[j,\star j] = j \wedge \star j + \star j \wedge j = 0$. Hence the equation of motion \eqn{jerw} of the
PCM is obeyed as well. Therefore we may use \eqn{jzolo} to demonstrate integrability of the action \eqn{nonabealdd} corresponding
to the non-Abelian dual of the PCM action \eqn{pcmmm}.

\section{A comparison in the $SU(2)$ case}

In this appendix we establish a precise relation between the background \eqn{dsbdn} and that in
\cite{Balog:1993es}. In the latter work the integrable model had a metric and antisymmetric tensor given by
\ba
&&ds^2 = {1\ov e^2} \left(dr^2 + {\b_0\ov x+1} ds^2(S^2)\right)\ ,
\nonumber\\
&& B = {1\ov e^2}{r-\a_0\ov x+1} {\rm Vol}(S^2)\ ,
\label{dsbd2n}
\ea
where
\be
r={\pi/2 - w\ov \sqrt{x^2-1}}\ , \quad \b_0
= {\cos^2 w \ov x+\cos2 w}\ ,\quad \a_0 = {\sqrt{x^2-1}\ov x+1} {\sin w\cos w\ov x+\cos 2w}\ ,\quad  x^2>1\ .
\ee
Letting
\be
w = {\pi \ov 2} +\a \ ,\qq {1\ov e^2 (x^2-1)} = {k (2 k+\k^2)\ov \k^2}\ ,\qq {2\ov x-1} = 4 {k(k+\k^2)\ov \k^4}\ ,
\ee
from which assuming $x>1$ we obtain
\be
x=1+{\k^4\ov 2 k(k+\k^2)}\ ,\qq e^2 = {4\ov \k^2} {k(k+\k^2)^2\ov (\k^2+2 k)^3}\ .
\label{kh99}
\ee
With these redefinitions \eqn{dsbd2n} and \eqn{dsbdn} become identical.
In addition, using \eqn{dkdjfh11} the expressions in \eqn{kjk90} become those in \eqn{kh99} (for $e^2$ we should take into
account the overall factor $k(1-\l^2)$).

\end{document}